\documentclass[aps,prl,amsmath,amssymbb,reprint,superscriptaddress]{revtex4-2}  %a tester format lettre

\usepackage{graphicx}% Include figure files
\usepackage{dcolumn}% Align table columns on decimal point
\usepackage{bm}% bold math
\usepackage[normalem]{ulem}

\usepackage[utf8]{inputenc}
\usepackage[T1]{fontenc}
\usepackage{graphicx}
\usepackage{enumitem}
\usepackage{amssymb}
\usepackage{amsthm}
\usepackage{cases}
\usepackage{tensor}
\usepackage{physics}
\usepackage{amsmath,amsfonts,amssymb}
\usepackage[dvipsnames]{xcolor}

\usepackage{hyperref}

\hypersetup{
   colorlinks,
    citecolor=blue,
    filecolor=black,
    linkcolor=blue,
    urlcolor=black
}

\definecolor{darkorange}{rgb}{1.0, 0.55, 0.0}

\newcommand{\bC}{\mathfrak{C}}
\newcommand{\colstrat}{\mathcal{S}}

\begin{document}

\title{Mean-field game approach to epidemic propagation on networks}

\newcommand{\majulab}{MajuLab, CNRS–UCA-SU-NUS-NTU International Joint Research Laboratory, 17543 Singapore}
\newcommand{\cqt}{Centre for Quantum Technologies, National University of Singapore, 17543 Singapore, Singapore}
\newcommand{\LPTMS}{Université Paris-Saclay, CNRS, LPTMS, 91405, Orsay, France}

\author{Louis Bremaud}
\affiliation{\LPTMS}
\author{Olivier Giraud}
\affiliation{\LPTMS}
\affiliation{\majulab}
\affiliation{\cqt}
\author{Denis Ullmo}
\affiliation{\LPTMS}

\date{\today}

\begin{abstract}
We investigate an SIR model of epidemic propagation on networks in the context of mean-field games. In a real epidemic, individuals adjust their behavior depending on the epidemic level and the impact it might have on them in the future. These individual behaviors in turn affect the epidemic dynamics. Mean-field games are a framework in which these retroaction effects can be captured. We derive dynamical equations for the epidemic quantities in terms of individual contact rates, and via mean-field approximations we obtain the Nash equilibrium associated with the minimization of a certain cost function. We first consider homogeneous networks, where all individuals have the same number of neighbors, and discuss how the individual behaviors are modified when that number is varied. We then investigate the case of a realistic heterogeneous network based on real data from a social contact network. Our results allow to assess the potential of such an approach for epidemic mitigation in real-world implementations. 
\end{abstract}
\maketitle

\subsection*{Introduction}

The lack of integration of dynamic human behavior into epidemic modeling remains a major limitation of contemporary epidemiological models \cite{arenas2020modeling, Covid_France_science, ferguson2020report} .  Indeed, individual behavior creates a time-dependent feedback on the transmission rate that is often out of reach for epidemiologists. Relevant human behavioral dynamics can be separated into two primary categories. The first corresponds to behaviors independent of epidemics, driven by routine patterns such as day/night cycles, weekdays versus weekends, holidays, and other habitual activities. The second category includes adaptive responses triggered by the epidemic itself, where individuals adopt precautionary behaviors such as using masks, avoiding handshakes, or reducing contact to lower infection risks \cite{poletti2010human}. These adaptive behaviors may arise spontaneously or be prompted by specific non-pharmaceutical interventions, creating a feedback loop that can significantly influence the epidemic's trajectory. Despite evidence of its importance \cite{ferguson2007capturing, epstein2009modelling}, particularly highlighted by the Covid-19 pandemic \cite{tang2022controlling}, this ``human-in-the-loop'' factor is often not considered in predictive models, where the dynamics of human behavior is treated instead as an external parameter \cite{Covid_France_science, guan2020modeling} acting on the transmission rate.

To address this limitation, theoretical approaches have been developed, including models that incorporate parallel information spread \cite{kiss2010impact, zhan2018coupling} or utilize payoff-based frameworks, as in Poletti's work \cite{poletti2009spontaneous}. In this study, we will focus on a recent and impactful approach: the Mean Field Game (MFG) paradigm. In short, MFGs are tools derived from game theory that enable to incorporate strategic interactions within systems involving a large number of agents. This game-theoretic framework makes it possible to account for anticipation effects arising from individuals optimizing intertemporal costs, and to describe ``free-rider'' behaviors, where individual optimization deviates from the collective societal optimum \cite{poletti2010human}. The solution associated with the MFG is referred to as a Nash equilibrium, meaning that no individual would benefit from modifying her strategy — that is, her behavior over the course of the epidemic — if the strategies of others remain unchanged. For a comprehensive mathematical introduction to MFG, see \cite{Lions_MFG}, and for applications of MFG to epidemiological modeling,  see \cite{Turinici_contact_rate_SIR_simple}, and \cite{guan2020modeling} for a recent review.

In this Letter we consider  the propagation of an epidemic where contacts between individuals can be described by a network. In such an instance, the structure of the underlying contact network, including factors such as contact heterogeneity, correlations, clustering, and other forms of network organization, has been demonstrated to have an important influence on epidemic dynamics \cite{watts1998collective, PhysRevE_cooperation_small_world_networks, Barthelemy2005_networks, dynamics_spreading_homogeneous_heterogeneous, Epidemc_spread_networks,newman2002spread}. For instance, heterogeneity is known to  significantly reduce the epidemic threshold on networks and to increase the propagation of the virus compared to a homogeneous network of the same average degree \cite{Barthelemy2005_networks, Epidemic_processes_complex_networks}. Correlations between degrees, reflected by the assortativity \cite{newman2003mixing} of the network and the clustering level \cite{newman2009random}, 
have also been shown to play a significant role in the propagation of epidemics.

On top of this network structure we implement a MFG framework. In the MFG approach, individuals are grouped into relevant classes to facilitate a mean-field treatment, requiring the identification of key factors driving individual behavioral responses. For instance in  \cite{bremaud2024mean} the age-based social structure is considered, along with the contact location (e.g. schools, households, workplaces), recognizing that age significantly influences the risk of infection in many diseases, while different locations lead to distinct contact patterns.  For epidemics on networks, we will make the basic assumption that individuals with the same number of neighbors behave in the same way. 

The objective of this paper is to examine how individuals' spontaneous behavioral responses are shaped by network structure within a Susceptible-Infected-Recovered (SIR) model on networks. We begin by presenting a model that is grounded in the MFG approach. We then analyze the impact of network degree by examining Nash equilibrium outcomes on homogeneous networks. Finally, we demonstrate how heterogeneity and network correlations give rise to specific effects on realistic networks.

\subsection*{The MFG framework on networks}

%\paragraph{Networks.}
We consider a population of $N$ individuals ($N$ large), represented by nodes of a network. The possible contacts of an individual are the neighboring nodes on the network. The number of these contacts is called the degree of the individual, denoted $k$. The degree distribution is denoted by $P(k)$, and the two-point degree correlation matrix is represented by $G_{kk'}=P(k'|k)$, which is the conditional probability for a given node of degree $k$ of having a neighbor of degree $k'$. Here we consider Markovian networks, defined by the fact that they are fully characterized by $P(k)$ and $G_{kk'}$ \cite{boguna2002epidemic}.

%\paragraph{SIR model.} 
Each individual, or node of the network, can be in one of three possible states $x = s, i, r$ for, respectively, susceptible, infected and recovered. 
Contamination occurs via edges connecting a susceptible individual to an infected individual.  The dynamics follows a standard Markov process : during the time interval $[t,t+dt[$, an edge between a susceptible and an infected individual transmits the disease with probability $\lambda(t) dt$. As in the basic SIR model, infected individuals recover from the disease during that time interval with probability $\gamma dt$. In view of the mean-field treatment of the problem, we assume that nodes of a given degree and state are equivalent, which allows us to characterize the dynamics by the average quantities $S_k,I_k,R_k$ giving the relative proportion of individuals of degree $k$ in the state susceptible, infected or recovered at time $t$. Moreover, we make the degree pairwise approximation \cite{compact_pairwise_heteregogeneous,keeling1999effects}, which posits that only correlations of degree and state between nearest neighbors on the network play a role in the dynamics. We thus introduce the conditional probability $G^{xy}_{kk'}$ for a given node to be of state $y$ and degree $k'$, knowing that this node has a neighbor of state $x$ and degree $k$, a quantity which accounts for all pairwise correlations inside the network. 

%\paragraph{Mean-field game setting.} 
On top of the above SIR model, we implement a MFG setting in which individuals control their own contact rate via a control variable $n(t)$ which they can adjust. We assume that the transmission rate between individuals $a$ and $b$ is symmetric and given by $\lambda^{(0)} n_a(t) n_b(t)$, where $\lambda^{(0)}$ represents the baseline rate in the absence of an epidemic. We make the assumption (see \cite{bremaud2024mean} for discussion) that those infected individuals who are responsible for contamination are asymptomatic (otherwise they would isolate themselves after becoming ill), and therefore behave as susceptible. Therefore, only the control variable of susceptible (or infected asymptomatic) individuals matters, since the others are taken out of the game. Physically, $n_a(t)$, which we call  the ``effort parameter'', represents the willingness of individual $a$ to engage in risky interactions with her neighbors. In the absence of effort we have $n_a(t)=1$, while the maximum effort corresponds to some fixed value $n_a(t) = \mathfrak{n}_{\textrm{min}}$.  In our mean-field framework, at the Nash equilibrium, the behavior of the agents only depends on their degree $k$, and one defines one control variable $n_k(t)$ for each degree. The effective transmission rate between individuals of degree $k$ and $k'$ is then given by $\lambda^{(0)} n_k(t) n_{k'}(t)$. Note that $n_k$ is assumed to be independent of the neighbor's degree $k'$. While this assumption may overlook some practical circumstances, it simplifies both the analytical and numerical resolution of the model.

\paragraph{Epidemic dynamics.}

 Considering now the dynamical equations describing our system, we introduce the following transition rates. 
 We denote by $T^k_{x \rightarrow z} dt$  the probability for the state $x$ of a node of degree $k$ to change to state $z$ in the time interval $dt$, and by  $T^{kk'}_{(x,y) \rightarrow (x',y')} dt$  the probability for an edge of type $(x, y)$ and degrees  $(k,k')$ to become of type $(x',y')$. As shown in the Supplemental Material \cite{supmat}, the only non-zero rates are 
\begin{subequations}\label{eq:transition_rates_networks}
\begin{align}
T_{i \rightarrow r }^k & = \gamma \; 
\label{eq:transition_rates_networksB}\\
T_{s \rightarrow i }^k & = \lambda^{(0)} n_k(t) \, k \sum_{k'} n_{k'}(t) G^{si}_{kk'}(t)  
\label{eq:transition_rates_networksA}\\
T^{kk'}_{(s,x) \rightarrow (i,x) }& \simeq \lambda^{(0)} n_k(t)  \Big[ n_{k'}(t) \delta_{x,i} + 
\label{eq:transition_rates_networksC}  \\
& \qquad (k-1)  \sum_{k''} n_{k''}(t) G^{si}_{kk''}(t)  \Big]  \, ,\notag
\end{align}
\end{subequations}
where in Eq.~\eqref{eq:transition_rates_networksC} we have used the pairwise approximation \cite{compact_pairwise_heteregogeneous,keeling1999effects}. The two terms in Eq.~\eqref{eq:transition_rates_networksC} reflect the fact that contamination of a susceptible node along a susceptible-infected edge can come from the infected neighbor (Kronecker delta) or from the $(k-1)$ other neighbors of the susceptible node.

With these notations, the SIR system for each degree can be expressed as
\begin{subequations}\label{eq:SIR_network_PA_MFG}
\begin{align}
\label{eq:SIR_network_PA_MFG1}
\dot{S}_k(t) &= -S_k(t) T^{k}_{s \rightarrow i } \\ 
% &= - \lambda^{(0)} n_k(t) S_k(t) k \sum_{k'} n_{k'}(t) G^{si}_{kk'}(t)\\
\dot{I}_k(t) &= S_k(t) T^{k}_{s \rightarrow i } - I_k(t) T^{k}_{i \rightarrow r } \\ 
% &= \lambda^{(0)} n_k(t) S_k(t) k \sum_{k'} n_{k'}(t) G^{si}_{kk'}(t) - \gamma I_k(t)
\dot{R}_k(t) &=  I_k(t) T^{k}_{i \rightarrow r } \;.
\label{eq:SIR_network_PA_MFG3}
\end{align}
\end{subequations}
Within the pairwise approximation \cite{compact_pairwise_heteregogeneous,keeling1999effects}, that is, neglecting three-point correlations (and beyond) which should appear in its evolution, the dynamics of $G^{si}_{kk'}$ is given (see Supplemental Material \cite{supmat}) by the coupled equations
\begin{equation}
\begin{aligned}
   \frac{d}{dt}(X_k G_{kk'}^{xy})  =  & 
   \sum_{x'y'} X'_k \, G_{kk'}^{x'y'} \, T^{kk'}_{(x',y') \to (x,y) } \\
 & \quad -  X_k \, G_{kk'}^{xy} \sum_{x'y'} T^{kk'}_{(x,y) \to (x',y') } 
   \; ,
   \label{eq:Gkk_final_time_evolution3_bis}
\end{aligned}
\end{equation}  
where $X_k$ denotes the relative proportion of agents of state $x$  in the class $k$.
The pairwise approximation has been shown to be very accurate on Markovian networks \cite{Wang_2017_unification}. 

The system \eqref{eq:transition_rates_networks}--\eqref{eq:Gkk_final_time_evolution3_bis} forms the Kolmogorov system of our MFG. Given the set $\colstrat=\{n_{k}(\cdot)\}_k$ of all collective strategies of degree-$k$ individuals at all times, this system describes the evolution of all epidemic rates.\\

\paragraph{Individual optimization}
In the MFG setting, the $n_k(t)$ are given as the result of individual optimization of agents and depend themselves on the epidemic rates. In order to obtain the $n_k(t)$, we assume that individuals of degree $k$ are sensitive to an intertemporal mean-field cost between the time $t$ at which the optimization is performed and the end of the game at time $T$ (assumed large). At time $t$, a representative {\em susceptible} individual $a$ of degree $k$ wishes to optimize the average cost \cite{bremaud2024analytical, bremaud2024mean}
\begin{equation}
\label{eq:final_cost_network}
  \bC\left(n_a(\cdot),\colstrat,t \right) = 
  \int_t^{T}   \left[\lambda_a(\tau) \ r_I + f_k \left (n_a(\tau)\right)\right] P_a(\tau|t)  d\tau \, ,
\end{equation}
in which we have introduced the force of infection perceived by individual $a$,
\begin{equation}
    \lambda_a(\tau) = \lambda^{(0)} n_a(\tau) k \sum_{k'} n_{k'}(\tau) G^{si}_{kk'}(\tau)\, ,
\end{equation} 
obtained in the same way as Eq.~\eqref{eq:transition_rates_networksA}, and $P_a(\tau|t) \equiv \exp \left[-\int^\tau_t \lambda_a(u) du\right]$ the probability for individual $a$ of still being susceptible at time $\tau>t$, knowing that she is susceptible at time $t$. In \eqref{eq:final_cost_network}, the cost function is the sum of a cost $r_I$, incurred in case of an infection, and a social cost $f_k$. Here we make the assumption that the infection cost $r_I$ is independent of $k$ (all individuals are equally affected by the disease), while the social cost of being deprived of contacts is likely to depend on the degree and hence is a function of  $k$.

From an individual's perspective, the best strategy at time $t$ is to tune her effort parameter $n_a(\tau), \tau>t$, in order to minimize her own foreseeable cost \eqref{eq:final_cost_network}. Introducing the value function
\begin{equation}
\label{eq:value_function_SIR_networks}
    U_a(t) = 
    \left\{
\begin{array}{ll}
\underset{n_a(\cdot)}{\min}\, \bC\left(n_a(\cdot),\colstrat,t \right),\,\, & a \textrm{ susceptible at }t\\
0,&a \textrm{ infected/recovered at }t,\\
\end{array}
\right.
\end{equation} 
one can show, following the same reasoning as in \cite{bremaud2024mean}, that 
\begin{equation}
\label{eq:HJB_SIR_network}
- \frac{dU_a}{dt} = \underset{n_a(t)}{\min}  \left[ \lambda_a(t) \left (r_I  - U_a(t) \right) + f_k(n_a(t))  \right]  \; .
\end{equation}
This is a differential equation for which the final condition $U_a(T) = 0$ is fixed; it is known as the Hamilton-Jacobi-Bellman equation of the game. Finally, the MFG setting requires a consistency condition to be at a Nash equilibrium, namely that the optimal strategy $n^*_a(t)$ which minimizes the right-hand side of \eqref{eq:HJB_SIR_network} should be the same as the one entering into the Kolmogorov system of equations  \eqref{eq:transition_rates_networks}--\eqref{eq:Gkk_final_time_evolution3_bis} for individuals with the same degree.  For any individual $a$ of degree $k$ one thus has
\begin{equation} \label{eq:Nash_SIR_networks}
n^*_a(t)= n_k(t) \; .
\end{equation} 
Equations  \eqref{eq:transition_rates_networks}--\eqref{eq:Gkk_final_time_evolution3_bis}, together with Eqs.~\eqref{eq:HJB_SIR_network} and \eqref{eq:Nash_SIR_networks}, form the MFG system of our game. We solve it numerically using a gradient descent approach (for details see \cite{bremaud2024these}).

\begin{table}
    \centering
     % \begin{tabular}{ c c c c c c c || c c c c c c c c c c | }
     %  \hline 
     % & & $(S_0,I_0,R_0)$ & & $\mu$ & & $\lambda^{(0)} \langle k \rangle$ & &  Interval $[\tilde{k}_i,\tilde{k}_{i+1}[$ & & & & &  \\ \hline
     %  & & (0.995, 0.005, 0)  & &  1/4 & & 4 & &  $\left( \ [2,5[,[5,7[,[7,10[,[10,19[,[19,100] \ \right)$ & & & & & \\ \hline 
     % & $r_I$ & $\mathfrak{n}_{\textrm{min}}$ & & $r$ & & $C$  & & Average $K_i$ & Distribution $\tilde{P}(K)$ & \\ \hline
     % &  50 & 0.1 & & 0.3 & &  0.4 & & (3.2, 5.4, 7.8, 12.5, 31.2) & (0.26, 0.25, 0.22, 0.20, 0.07) & \\ \hline
     % \end{tabular}\\
            \begin{tabular}{ | c |}
     \hline 
     $(S_0,I_0,R_0)=(0.995, 0.005, 0),\quad \gamma=1,\quad\lambda^{(0)} \langle k \rangle =4$  \\ \hline
   $r_I=50,\quad\mathfrak{n}_{\textrm{min}}=0.1$ \\
   \hline
    \end{tabular}
    \caption{Table of parameters used in our simulations. $\lambda^{(0)} = 4/ \langle k \rangle$ allows to compare appropriately epidemics on different networks by rescaling the infection rate and keep a constant infection probability $\lambda^{(0)} \langle k \rangle$  on average.}
    \label{table:parameters}
\end{table}

For all our simulations, the parameters characterizing the epidemics are the ones given in Table \ref{table:parameters}. For the social cost function, we chose the specific form
\begin{equation}
    f_k^{\epsilon}(n(t)) = k^{\epsilon} \left(\frac{1}{n(t)}-1 \right)\;,  \quad \epsilon = 0,1 \;, 
\end{equation}
which allows us to explore different regimes of social dependence to neighbors. Physically, the choice $\epsilon = 1$ implies that a constant social cost of $(\frac{1}{n_a(t)} - 1)$ is assigned to each neighbor, which means that for a fixed fraction of contacts lost, an individual with a higher number of neighbors is more impacted than an individual with fewer neighbors. In the case $\epsilon = 0$ the social cost is the same for all individuals, whatever their degree. 
%Finally, the case $\epsilon = -1$ corresponds to individuals who feel less close to their neighbors when their contacts increase. Although this dependence may seem somewhat counter intuitive compared to the previous choices,  it may correspond to the situation where a quest for a large network is associated with a lack of interest in maintaining deep bonds with some close neighbors (friends or family).

%%%%%%%%%%%%%%%%%%%%%%%%%%%%%%%%%%%%%%%%%%%%%%%%%%%%%%%%%%%%%%%%%%%%%%%%%%%%%%%%%%%%%%%%%%%%%%%%%%%%%%%%%%%%%%%%%%%%%%%%%%%%%%%%%%%%%%%%%
%%%%%%%%%%%%%%%%%%%%%%%%%%%%%%%%%%%%%%%%%%%%%%%%%%%%%%%%%%%%%%%%%%%%%%%%%%%%%%%%%%%%%%%%%%%%%%%%%%%%%%%%%%%%%%%%%%%%%%%%%%%%%%%%%%%%%%%%%
\subsection*{Homogeneous networks}
\begin{figure}[!t]
    \centering
    \includegraphics[width=\linewidth]{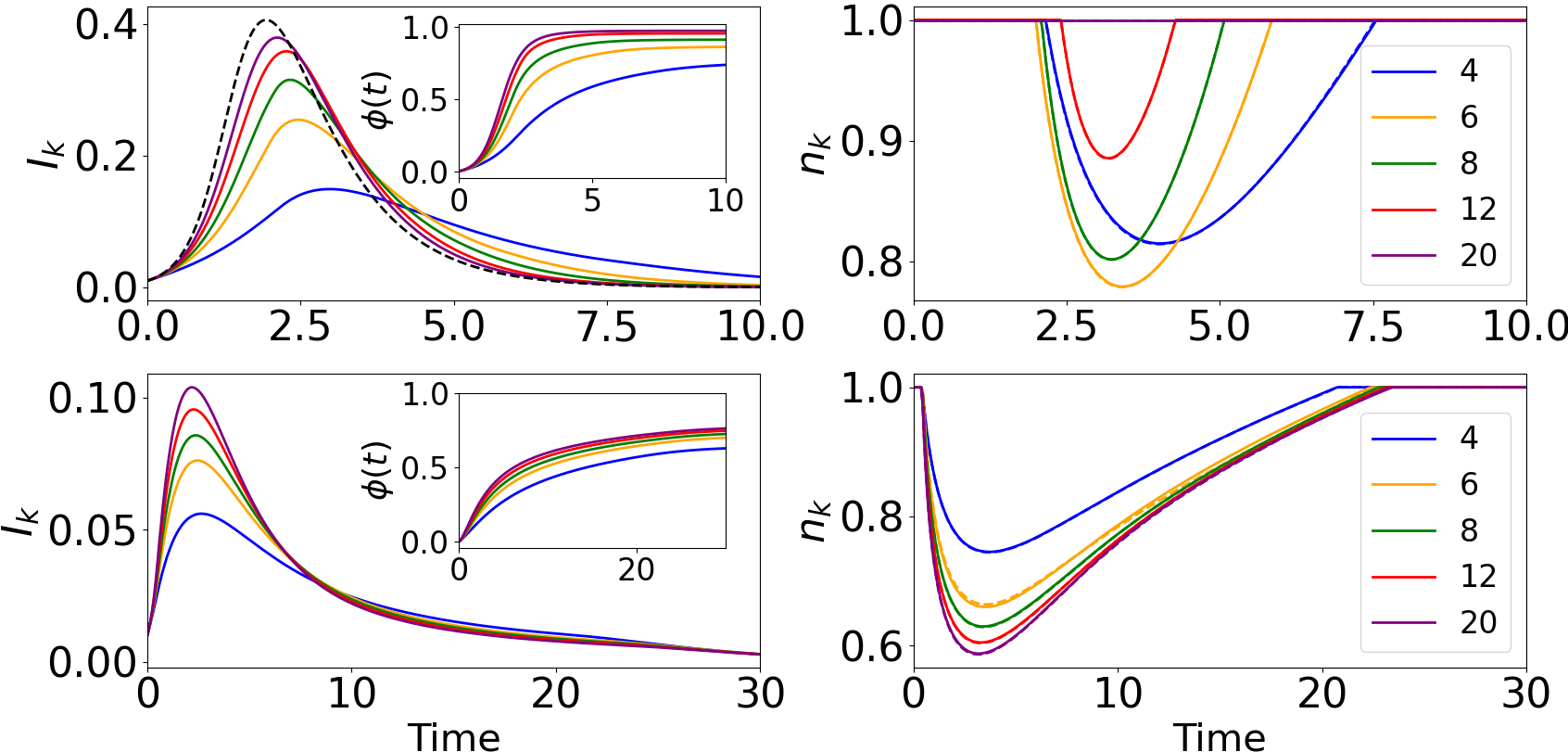}
    \caption{Left column: Dynamics of infected individuals, corresponding to the Nash equilibrium, with the parameters of Table~\ref{table:parameters} for different homogeneous networks, with $k=4$ (blue), $6$ (orange), $8$ (green), $12$ (red), $20$ (purple) and classical SIR model (black dashed); the social cost function is $f_k^{\epsilon}$ with $\epsilon= 1$ (top)  and $\epsilon=  0$ (bottom). Inset: dynamics of the probability $\phi(t) = 1 - P(t|0)$ to be infected before $t$. Right column: Dynamics of the corresponding individual effort parameter, with the same parameters and color code as for the left column.}
    \label{fig:Nash_homogeneous}
\end{figure}

%%%%%%%%%%%%%%%%%%%%%%%%%%%%%%%%%%%%%%%%%%%%%%%%%%%%%%%%%%%%%%%%%%%%%%%%%%%%%%%%%%%%%%%%%%%%%%%%%%%%%%%%%%%%%%%%%%%%%%%%%%%%%%%%%%%%%%%%%
%%%%%%%%%%%%%%%%%%%%%%%%%%%%%%%%%%%%%%%%%%%%%%%%%%%%%%%%%%%%%%%%%%%%%%%%%%%%%%%%%%%%%%%%%%%%%%%%%%%%%%%%%%%%%%%%%%%%%%%%%%%%%%%%%%%%%%%%%
We first consider the simplest case of homogeneous networks (or regular graphs), where each node has the same number $k$ of neighbors. After numerically solving the system of equations discussed above and reaching a Nash equilibrium, we obtain the epidemic rates and associated effort parameters. They are displayed in Fig.~\ref{fig:Nash_homogeneous} for the two different possibilities $f_k^{0,1}$. Several observations can be made. 

First, we observe in Fig.~\ref{fig:Nash_homogeneous} that while individuals reduce their contact rate predominantly during the epidemic peak, their maximal effort occurs slightly after the peak is reached (see, for instance, the case $k=4$ on the first row), and they maintain their effort well beyond the peak. This suggests that individuals engage in a form of ``reverse anticipation''.  More precisely, it is not the anticipation of the incoming epidemic that motivates their behavior, but the compound effect of the actual (present time) intensity of the epidemic and of the {\em anticipation of its end}.  Indeed, at the onset of the epidemic, the prospect of maintaining a significant effort for the whole duration of the epidemic, while the latter is still growing slowly and individuals anticipate that collective immunity will not be reached anytime soon, appears more  costly (with our choice of parameters) than paying the ``one time'' cost of infection.
However, as collective immunity is in sight, shortly before the epidemic peak and for some time after, it becomes advantageous to make efforts to avoid infection, since the epidemic is still severe, and the remaining time before the epidemics is over is reasonably short.
It then becomes advantageous for susceptible individuals to make significant efforts, as they have a good chance of avoiding infection forever if they protect themselves for a relatively short period. 

While the mechanism described above is rather generic, the precise range and intensity at which it is at play of course depends on the choice of parameters.  In particular, epidemics on random homogeneous networks progress faster and are more intense as $k$ increases \cite{bremaud2024analytical}.
%\sout{In the case of constant  $f_k$ ($\epsilon=0$, right column of Fig.~\ref{fig:Nash_homogeneous}),  this results in  individuals maintaining their effort a significant time after the epidemic peak.  This effort furthermore increases in duration and in intensity with $k$.  Here, although collective immunity has largely been achieved, individuals with large $k$ are still at significant risk of becoming infected if they do not maintain significant effort until the epidemic has completely disappeared; and since in the  $\epsilon = 0$ case, this effort is not too costly,  the agents accept to bear it.}
For constant $f_k$ ($\epsilon=0$, second row of Fig.~\ref{fig:Nash_homogeneous}), the ratios between social effort and infection cost remain essentially constant across degrees, and are fairly low for our choice of parameters.  This leads to effort patterns that are similar across degrees, with individuals tending to protect themselves by ``flattening'' the infection curve $\phi(t)$, thereby minimizing their probability of infection. The only difference between classes is that individuals with higher degrees face more intense epidemics, requiring greater and more prolonged effort while maintaining the same overall pattern.
On the other hand, when the social cost $f_k$ increases with $k$ ($\epsilon=1$, first row of Fig.~\ref{fig:Nash_homogeneous}), this increasing social cost may compete with the one of the infection. As Fig.~\ref{fig:Nash_homogeneous} shows,  these two  factors essentially balance each other around  a critical value $k^* \simeq 6$, leading there to a significant intensity of efforts. However, below this threshold, the epidemic is not sufficiently virulent, and  above $k^*$  efforts becomes too costly  to justify  a strong reduction of social contact. As $k \to \infty$, individual behavior converges to the effortless parameter $n(t) = 1$, and the infection curve approaches that of the classical SIR model (see dashed curve in Fig.~\ref{fig:Nash_homogeneous}).

%%%%%%%%%%%%%%%%%%%%%%%%%%%%%%%%%%%%%%%%%%%%%%%%%%%%%%%%%%%%%%%%%%%%%%%%%%%%%%%%%%%%%%%%%%%%%%%%%%%%%%%%%%%%%%%%%%%%%%%%%%%%%%%%%%%%%%%%%
%%%%%%%%%%%%%%%%%%%%%%%%%%%%%%%%%%%%%%%%%%%%%%%%%%%%%%%%%%%%%%%%%%%%%%%%%%%%%%%%%%%%%%%%%%%%%%%%%%%%%%%%%%%%%%%%%%%%%%%%%%%%%%%%%%%%%%%%%
\subsection*{Heterogeneous networks}
%%%%%%%%%%%%%%%%%%%%%%%%%%%%%%%%%%%%%%%%%%%%%%%%%%%%%%%%%%%%%%%%%%%%%%%%%%%%%%%%%%%%%%%%%%%%%%%%%%%%%%%%%%%%%%%%%%%%%%%%%%%%%%%%%%%%%%%%%
%%%%%%%%%%%%%%%%%%%%%%%%%%%%%%%%%%%%%%%%%%%%%%%%%%%%%%%%%%%%%%%%%%%%%%%%%%%%%%%%%%%%%%%%%%%%%%%%%%%%%%%%%%%%%%%%%%%%%%%%%%%%%%%%%%%%%%%%%

\begin{table}
    \centering
            \begin{tabular}{ | c |}
     \hline 
      Intervals $[\tilde{k}_i,\tilde{k}_{i+1}[=[2,5[,[5,7[,[7,10[,[10,19[,[19,100]$  \\ \hline
    Average $K_i=(3.2, 5.4, 7.8, 12.5, 31.2)$ \\ \hline
    Distribution $\tilde{P}(K)=(0.26, 0.25, 0.22, 0.20, 0.07) $ \\ \hline
    $ G_{KK'} = \begin{pmatrix}
    0.76 & 0.03 & 0.04 & 0.06 & 0.11 \\
    0.02 & 0.78 & 0.04 & 0.06 & 0.10 \\
    0.02 & 0.03 & 0.79 & 0.06 & 0.10 \\
    0.02 & 0.03 & 0.04 & 0.80 & 0.11 \\
    0.03 & 0.06 & 0.07 & 0.11 & 0.72 \\
    \end{pmatrix}$  \\ \hline
    \end{tabular}
    \caption{Parameters characterizing the realistic heterogeneous network used for Fig.~\ref{fig:Nash_heterogeneous}: the $5$ batches $[\tilde{k}_i,\tilde{k}_{i+1}[$, the average degree $K_i$ of the nodes in each interval, and the corresponding degree distribution $\tilde{P}(K)$ and correlation matrix $G_{KK'}$.}
    \label{table:parameters-heterogeneous}
\end{table}

We now investigate the more realistic case of a heterogeneous network. SIR model on such networks is usually studied by considering a scale-free distribution $P(k)$ \cite{Epidemic_processes_complex_networks}. As the correlation matrix $G_{kk'}$ plays a crucial role in the MFG equations, we choose to investigate a realistic network constructed in the following way: 
%\paragraph{Construction of generic realistic networks.} 
We build $P(k)$ based on the work of Eubank \textit{et al.}~\cite{eubank2004modelling} and Béraud \textit{et al.}~\cite{beraud2015french}. We define it as a piecewise power-law distribution $P(k) \propto k^{\eta(k)}$ with  $\eta(k) = 1$ for $k \in [2,5], -1.5$ for $k \in [5,10], -3$ for $k \in [10,100]$, which gives a maximum of around 5 contacts per day. We chose the above exponents $\eta(k)$ and intervals for $k$ in such a way that the range of $k$, average, standard deviation and  maximum of that distribution are consistent with \cite{ beraud2015french}. In order to perform the numerical simulations in a reasonable time, we split our distribution $P(k)$ into batches containing approximately the same number of nodes. Namely, we consider that all nodes with degree $k\in [\tilde{k}_i,\tilde{k}_{i+1}[$ can be treated as nodes with degree $K_i$, with $K_i$ the average degree of the nodes in that interval. Our choice for the batches is given in Table~\ref{table:parameters-heterogeneous}. The quality of this approximation is demonstrated in Section II of the Supplemental Material \cite{supmat}.

For a given correlation matrix $G_{kk'}$, one can introduce an assortativity coefficient $r\in [-1,1]$, defined precisely in \cite{newman2003mixing}. A  positive $r$ intuitively means that high-degree individuals will tend to have contacts with high-degree individuals, and similarly for  low-degree individuals.   Social contact networks are known to be assortative, and here  we choose $r$ approximately equal to $0.3$, compatible with the kind of networks described in \cite{newman2003mixing}. Using the Newman rewiring algorithm \cite{di2024generation}, we obtain a matrix $G_{kk'}$ averaged over $10$ networks of $20 000$ nodes with $r \simeq 0.3$.

\begin{figure}[ht]
    \centering
    \includegraphics[width=\linewidth]{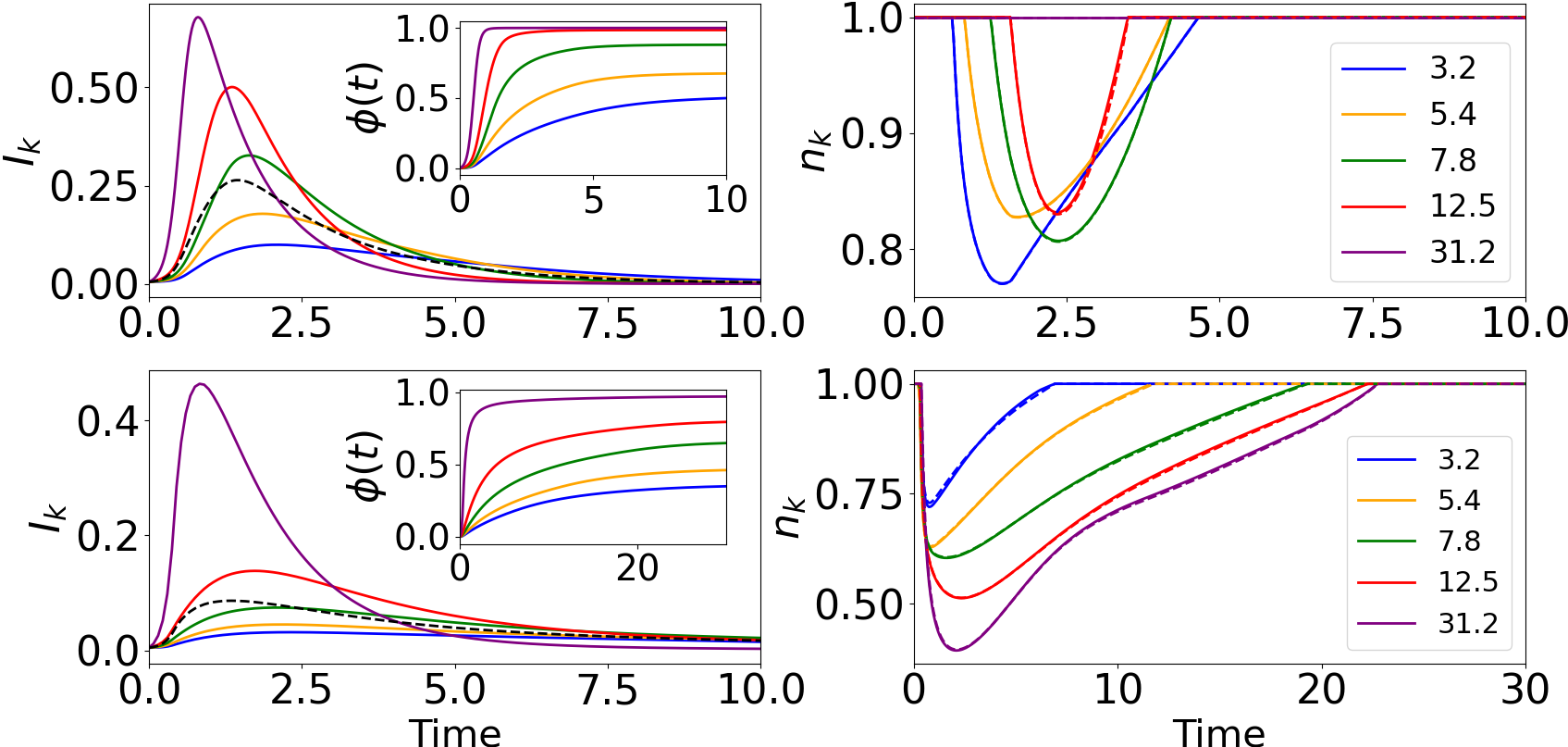}
    \caption{Left panels: Dynamics of infected individuals at Nash equilibrium for different batches, with the parameters of Tables~\ref{table:parameters} and \ref{table:parameters-heterogeneous}. Inset: dynamics of the probability $\phi(t)$ to be infected before $t$. Right panels: Dynamics of the corresponding individual effort parameter.  Colored solid lines corresponds to the dynamics (for infected and effort parameter) associated with each batch of the network:  $K=3.4$ (blue), $5.4$ (orange), $7.8$ (green), $12.5$ (red), $31.2$ (purple). Each row represents a specific choice of $f_k^{\epsilon}$: $\epsilon= 1, 0$ for the first and second row, respectively.}
    \label{fig:Nash_heterogeneous}
\end{figure}

%\paragraph{Nash equilibrium.} 
The dynamics of the epidemics and the associated effort parameters at the Nash equilibrium are obtained by solving Eqs.~\eqref{eq:transition_rates_networks}--\eqref{eq:Nash_SIR_networks}.  We assume that $G^{xy}_{kk'}(0) = X_k(0) G_{kk'}$, which indicates that there is no correlation between states and degrees at time $t=0$. The results are displayed in Fig.~\ref{fig:Nash_heterogeneous} for the two different choices of $f_k^{\epsilon}$. The specific impact of a realistic distribution, together with the interactions between classes (heterogeneity), can be captured. In all cases, we observe that, contrary to what might  be expected, the spread, as a function of $k$, of the total number of infected at $T$ (inset panel) increases compared to the homogeneous case. This is related to the collective immunity that is now achieved at the network level (and not for each degree class as in the homogeneous case). This essentially means that very high-degree individuals cannot really avoid the disease, since they are infected before all other classes. For them, applying a strong social distancing would only delay the infection peak, but would not lead to heard immunity. Then, the epidemic continues to spread in the network even though all high-degree individuals have been infected, since they represent a very small fraction of all nodes. On the other hand, low degree individuals take advantage of this situation and reach a collective immunity with a rate $I_k$ below that required in the homogeneous case.  In fact, more than the proportion of infected individuals among high-degree individuals, the average degree of the remaining susceptible nodes decreases rapidly, which helps achieve herd immunity. 

Differences in infection rates result in infection curves that strongly depend on the degree. For $\epsilon=1$ (Fig.~\ref{fig:Nash_heterogeneous}, upper right panel), interactions between classes influence the competition between costs in a complex manner: the curve tails shorten with increasing degree, while effort levels decrease non-monotonically. In contrast, for $\epsilon=0$ (Fig.~\ref{fig:Nash_heterogeneous}, lower right panel), effort patterns become degree-specific in a more understandable way: high-degree individuals protect themselves, while low-degree individuals benefit from the collective immunity achieved by others more rapidly.

\subsection*{Conclusion}
In the present Letter we studied the problem of epidemic propagation on networks from the point of view of mean-field games. This allowed us to analyze how individual behavior may affect the outcome of an epidemic when that behavior itself is modified at each time by the epidemic. In our model, individuals can tune the intensity of the contacts they are willing to have with others (effort parameter)
 in order to optimize the cost that this choice will make them incur in the future. We showed that this interplay can be described by a Hamilton-Jacobi-Bellman system of equations for the individual costs and effort parameters, coupled with a set of Kolmogorov equations describing the epidemic dynamics.

Our MFG approach to networks highlights the ``reverse anticipation'' effect, where individuals adjust their behavior in anticipation of the end of the epidemic - a phenomenon likely to be observed in contexts other than networks. This anticipation can be brief, as in the case of increasing social costs with $k$, or have a long tail, as in the case of constant social costs, when efforts effectively reduce the probability of infection without being too costly.   In the homogeneous case with $\epsilon = 1$, the model shows a balance between the increasing social cost with $k$ and the higher epidemic costs experienced by individuals with high degrees, while a more homogeneous behavior is observed at $\epsilon = 0$.  The introduction of heterogeneity and assortativity in a realistic network leads to differentiated collective immunity at the node level: low-degree individuals benefit from the fast spreading of the epidemic among high-degree individuals, which reduces the effective connectivity of the remaining susceptible network. Contrary to expectations, heterogeneity reduces costs for low-degree individuals, while positive assortativity weakens this protection, as it tends to reduce heterogeneity between classes.

In both cases, the role of the social cost $f$ on the behavior of individuals is crucial, even though the only variations of  $f$  we considered were the ones associated with its  degree $k$. Our work underlines that a precise description of the behavior of $f$ is a key element to go further in the practical implementation of MFG frameworks. This endeavor should benefit from the fact that the social cost properties should show little variation across epidemics, allowing large surveys to obtain the dependencies of $f$.

\bibliography{biblio_file}

\end{document}

\title{Supplemental Material:\\ Mean-field game approach to epidemic propagation on networks}

\author{Louis Bremaud}
\author{Olivier Giraud}
\author{Denis Ullmo}

\maketitle

\section{Derivation of Eq.~(1)-(3):  dynamics within the pairwise approximation}
\label{appA:pairwise-approx}

\subsection{Setting and definitions}
\label{settings_defs}
In this Section, we provide a derivation of Eqs.(1)-(3) of the main text.  
% To lighten notations, we assume that the transmission rate 
% \begin{equation}
%     \lambda_{kk'}(t) = \lambda(0) n(k) n(k') \label{SI:lambdakkp}
% \end{equation}is independent of $k,k'$, and will re-inject this dependence only at the end.  
Nodes are labeled with Greek letters ($\alpha,\beta,\cdots$), and states $\{s,i,r\}$ with Roman letters ($x,y,x,\cdots)$. We denote by $d_\alpha$ the degree of node $\alpha$, and by $A_{\alpha\beta}$ the adjacency matrix ($A_{\alpha\beta}=1$ if there is an edge between $\alpha$ and $\beta$, and 0 otherwise). The state $x\in \{ s,i,r\}$ of the node $\alpha$ at time $t$ is denoted $c_t(\alpha)$; more generally $c_t(\alpha_1, \alpha_2,...)=(x_1,x_2,...)$ denotes the states $x_i$ of nodes $\alpha_i$ at time $t$. 

%will use $\backslash mathbb$  for sets.  For instance $\sV_\alpha = \{\beta \,/\, A_{\alpha\beta} \smeq 1\}$  is the set of all neighbors of  $\alpha$.  The degree $d_\alpha$ of $\alpha$ is $\#\sV_\alpha$, where $\#\sE$ the cardinal of set $\sE$.

We introduce various sets characterizing the network:
\[
\begin{array}{|c|l|l|}
\hline
\textbf{Symbol} & \textbf{Definition} & \textbf{Description} \\
\hline
\sD_k & \{ \alpha \;/\; d_\alpha \smeq k \} & \text{Nodes of degree } k \\
\hline
\sV_\alpha & \{ \beta \;/\; A_{\alpha\beta} \smeq 1 \} & \text{Neighbors of node } \alpha\\
\hline
\sTD_k & \{ (\alpha,\beta) \;/\; \alpha \in \sD_k \;\&\; \beta \in \sV_\alpha \} & \text{Oriented edges starting from a node of degree } k \\
\hline
\sTD_k^x & \{ (\alpha,\beta) \;/\; \alpha \in \sD_k \;\&\; \beta \in \sV_\alpha \} & \text{Oriented edges starting from a node of degree } k \\
&&\text{ and state }x\\
\hline
\sE_{kk'} & \{ (\alpha,\beta) \;/\; \alpha \in \sD_k \;\&\; \beta \in \sD_{k'} \;\&\; A_{\alpha\beta} \smeq 1 \} & \text{Oriented edges between nodes of degree } k \text{ and } k' \\
\hline
\sWaux %\sW{k}{k'}{k''} 
& \{ (\alpha,\beta,\gamma) \in \sD_k \times \sD_{k'} \times \sD_{k''} \;/\; \beta,\gamma \in V_\alpha \} & \text{Oriented wedges of degrees } (k, k', k'')\\
\hline
\end{array}
\]
Note that in the definition of $\sWaux$, the two edges are oriented but also the wedge itself, i.e.~the ordering of the two edges matters.

%"Could you transform this into a table:"
% \begin{align*}
%     \sD_k & = \{ \alpha \;/\; d_\alpha \smeq k \} 
%         \qquad  \mbox{(nodes of degree $k$)} \\
%     \sV_\alpha &= \{ \beta \; / \; A_{\alpha\beta} = 1 \} 
%         \qquad  \mbox{(neighbors of node $\alpha$)}  \\
%     \sTD_k    &= \{ (\alpha,\beta)  \;/\; \alpha \in \sD_k \;\&\; \beta \in \sV_\alpha \} \qquad \mbox{  (oriented edges starting from a node of degree $k$)} \\
%     \sE_{kk'} &= \{ (\alpha,\beta)  \;/\; \alpha \in \sD_k \;\&\; \beta \in \sD_{k'} \;\&\; A_{\alpha\beta} \smeq 1 \} \qquad \mbox{ (oriented edges between nodes of degree $k$  and $k'$ ).} 
% \end{align*}
We furthermore introduce sets that characterize the epidemic status of nodes on the network:
\[
\begin{array}{|c|l|l|}
\hline
\textbf{Symbol} & \textbf{Definition} & \textbf{Description} \\
\hline
\sG^x_k & \{ \alpha \in \sD_k \; /\; c_t(\alpha) \smeq x \} & \text{Nodes of degree } k \text{ and state } x \\
\hline
\sG^{xy}_{kk'} & \{ (\alpha,\beta) \in \sE_{kk'} \;/\; c_t(\alpha,\beta) = (x,y) \} & \text{Oriented edges linking a node of degree } k \\
& & \text{and state } x \text{ to a node of degree } k' \text{ and state } y \\
\hline
\sG_{kk'k''}^{xyz} & \{ (\alpha,\beta,\gamma) \in \sWaux \;/\; c_t(\alpha,\beta,\gamma) = (x,y,z) \} & \text{Wedges with given status and degree} \\
\hline
\end{array}
\]
% \begin{align*}
%   \sG^x_k & = \{ \alpha \in \sD_k  \; /\; c_t(\alpha) \smeq x \}  \qquad \mbox{(nodes of degree $k$ and state $x$)} \\
%     \sG^{xy}_{kk'}  &= \{ (\alpha,\beta) \in \sE_{kk'} \;/\; c_t(\alpha,\beta) = (x,y) \} \qquad \mbox{(oriented edge linking a node of degree $k$ and state $x$ to a node of degree $k'$ and state $y$)}\\
% \sW{k}{k'}{k''}& = \{ (\alpha;\beta,\gamma)  \in \sD_k\times \sD_{k'} \times \sD_{k''} \;/\; (\beta,\gamma) \in V_\alpha^2 \} \qquad \mbox{(wedges  of degrees  $(k; k', k'')$)}\\
% \sG_{kk'k''}^{xyz}& = \{ (\alpha,\beta,\gamma) \in \sW{k}{k'}{k''} \; /\; 
% c_t(\alpha,\beta,\gamma) = (x,y,z) \} \qquad \mbox{(wedges with given status and degree)}
% \end{align*}

The number of elements of a set $\sS$ will be denoted by $\#\sS$. In particular, we have $\#\sD_k=N_k$, $\#\sTD_k = k N_k$, and $\#\sTD_k^x = k N_k X_k$.

\subsection{Correlation matrices and transition rates}

\subsubsection{One-point, two-point and three-point correlators}
With the above notation, the relative proportions $X_k=S_k,I_k,R_k$ of individuals of degree $k$ in the state susceptible, infected or recovered is given by
\begin{equation}
X_k = \frac{\#\sG^x_k}{N_k},
\end{equation}
which is a one-point correlator.

In the same way the two-point correlations between adjacent nodes are given by
\begin{equation}
 G^{xy}_{kk'}   \equiv \frac{\# \sG^{xy}_{kk'}}{ \# \sTD_k^x} = \frac{\# \sG^{xy}_{kk'}}{ k N_k X_k} \; . \label{eq:Gdef} 
 \end{equation}
By symmetry $(\# \sG^{xy}_{kk'}) = (\# \sG^{yx}_{k'k})$, and thus we recover the detailed balance condition $k N_k X_k G^{xy}_{kk'} = k' N_{k'} Y_{k'} G^{yx}_{k'k} $.

Finally, the three points correlation matrix can be derived from the number of elements of the set $\sG_{kk'k''}^{xyz}$, which we shall discuss in Section \ref{pairwise} below.

\subsubsection{Bare transition rates}
The two processes that can lead to a transition between states on the network are processes transforming nodes and processes transforming edges. In the SIR model, during a time interval $[t,t+dt[$, a node has some probability to go from state $i$ to $r$, and an edge connecting a node $s$ and a node $i$ has some probability to become an edge connecting $i$ to $i$.

Consider a node $\alpha$ in the state $x$. We  note $\Lambda_{x}^{x'}$  the rate of transformation of $\alpha$ at time $t$ from state $x$ to state $x'$ {\em which is not due to any interaction with its neighbors}. For the SIR model, the only such bare process corresponds to the recovery process, where a node $i$ transforms into $r$, and thus the only nonzero rate is
\begin{equation}
    \Lambda_{i}^{r} =  \gamma  \; .
\end{equation}
Similarly, consider an undirected edge between node $\alpha$ of state $x$ and degree $k$ and node $\beta$ of state $y$ and degree $k'$. We  denote $\Lambda_{xy}^{x'y'}$  the rate of transformation at time $t$ from state $xy$ to state $x'y'$ which is due {\em only to the interaction between $\alpha$  and $\beta$}. For the SIR model, the only such bare process corresponds to  an edge $si$ transforming into $ii$. In the case of the SIR on networks considered in the main text, the only non-zero rate reads
\begin{equation}
    \Lambda_{si}^{ii}= \lambda_{kk'} (t)  = \lambda(0) n_k n_{k'} \; .
\end{equation}
There is an important distinction to make between directed and undirected edges. The sets defined in Section \ref{settings_defs} count directed edges; this accounts for the fact that a node whose state is modified by the transformation of an edge can potentially sit at either ends of that edge. However,
the above rates apply to undirected edges. Note also that the above rates depend on $k$ and $k'$; to ease the reading we have omitted this dependency in the notation. 
%\rem{ [Denis] Oui, En fait il faudrait rajouter les indices $k.k'$ pour les $\Lambda$'s.  Mais c'est sans doute un peu lourd.  Il faudrait peut être juste le préciser en mots.} 

\subsubsection{Dressed transition rate for the one-point correlators}
In order to determine the time evolution of the correlators, we need to calculate their rate of change taking into account all possible processes (single-node events, events involving nearest neighbors, and so on), which we call {\it dressed} transition rates.

Let $\sT^{k}_{(x;t) \to (x';t+dt) }=\left\{ \alpha\in \sD_{k} \; / \; 
    c_t(\alpha) \smeq (x) \, \& \, c_{t+dt}(\alpha) \smeq (x') 
    \right\}$ be the set of nodes of degree $k$ that change their state from $x$ to $x'$ during the time interval $[t,t+dt[$. The corresponding transition rate $T^{k}_{x \to x' }$, defined by
\begin{equation} \label{SI:drate-1p}
  \# \sT^{k}_{(x;t)  \to (x';t+dt) } 
   =  \left(\# \sG^{x}_{k} \right) T^{k}_{x \to x' } dt \; ,
\end{equation}
gives the transition probability of a node of degree $k $ from $x$ to $x'$ during $dt$.
Neglecting contributions of order $(dt)^2$ (i.e.\ the probability of two independent events happening during the interval $dt$), a node is modified either by a single-node event (for SIR, the node transition $i \to r$), or by the transition of an indicent edge (for SIR, the edge transition $si\to ii$). 
Summing the corresponding probabilities, we have 
\begin{equation} 
\label{tkxt}
   \left(\# \sT^{k}_{(x;t)  \to (x';t+dt) } \right)
   =  \left(\# \sG^{x}_{k} \right)\Lambda_x^{x'} dt + \sum_{k';y,y'} \left(\# \sG^{xy}_{kk'} \right) 
   \Lambda_{xy}^{x'y'} dt \; ,
\end{equation}
Dividing both sides of \eqref{tkxt} by $(\#\sG^x_k) dt$ and using Eq.~\eqref{eq:Gdef}, we get
\begin{equation} 
T^{k}_{x \to x'}   
   =  \Lambda_x^{x'}  +  k \sum_{k';y,y'}  G^{xy}_{kk'}  \Lambda_{xy}^{x'y'} \; .
\end{equation}
For our SIR model on networks, this reduces to $T^{k}_{i \to r} = \Lambda_i^{r} = \gamma $, and $T^{k}_{s \to i} =  k \sum_{k'} \lambda_{kk'}(t) G^{si}_{kk'}$, which is exactly Eqs.~(1a)--(1b) of the main text.

\subsubsection{Dressed transition rates for the two-point correlator}

To describe the time evolution of the $G^{xy}_{kk'}$, we need to additionally take into account the probability that an edge between two nodes $\alpha$ and $\beta$ change due to processes that involves nodes other than $\alpha$ or $\beta$, but to which $\alpha$ or $\beta$ are linked. We thus introduce the sets of directed edges going from state $xy$ to state $x'y'$ between $t$ and $t+dt$,
\begin{equation}
    \sT^{kk'}_{(x,y;t) \to (x',y';t+dt) } \equiv \left\{ (\alpha,\beta)\in \sE_{kk'} \; / \; 
    c_t(\alpha,\beta) \smeq (x,y) \, \& \, c_{t+dt}(\alpha,\beta) \smeq (x',y') 
    \right\}
\end{equation}
and we define the corresponding dressed transition rates $T^{kk'}_{(x,y) \to (x',y') }$ as
\begin{equation} \label{SI:drate-2p}
   \left(\# \sT^{kk'}_{(x,y;t)  \to (x',y';t+dt) } \right)
   =  \left(\# \sG^{xy}_{kk'} \right) T^{kk'}_{(x,y) \to (x',y') } dt \; .
\end{equation}
Neglecting again terms of order $(dt)^2$ (i.e.\ the probability of two independent simultaneous processes), we get the following contributions:
\begin{enumerate}
    \item $(x\smneq x')$ and $(y\smneq y'$). The only process transforming two connected nodes is the process transforming the edge that connects them:
    \begin{equation} 
    \left(\# \sT^{kk'}_{(x,y;t) \to (x',y';t+dt)} \right)  =  \left(\# \sG^{xy}_{kk'} \right) \Lambda_{xy}^{x'y'} dt.
    \end{equation}
    \item $(x\smneq x')$ and $(y \smeq y'$). The processes involved are the transformation of the edge $\alpha-\beta$ or of the node $\alpha$ alone, as well as the transformation of an edge connecting $\alpha$ with any of its other neighbors:
    \begin{equation} \label{eq:sT2}
    \left(\# \sT^{kk'}_{(x,y;t) \to (x',y;t+dt) } \right)  =  \left(\# \sG^{xy}_{kk'} \right) \left[\Lambda_{xy}^{x'y} + \Lambda_{x}^{x'} \right] dt
    +
    \sum_{k''zz'} \left(\# \sG_{kk'k''}^{xyz}\right) \Lambda_{xz}^{x'z'} dt
    \end{equation}
    \item $(x\smeq x')$ and $(y \smneq y'$). Symmetrically, as in \eqref{eq:sT2} we have
    \begin{equation} \label{eq:sT3}
    \left(\# \sT^{kk'}_{(x,y;t) \to (x,y';t+dt) } \right)  =  \left(\# \sG^{xy}_{kk'} \right) \left[\Lambda_{xy}^{xy'} + \Lambda_{y}^{y'} \right] dt
    +
    \sum_{k''zz'} \left(\# \sG_{k'kk''}^{yxz}\right) \Lambda_{yz}^{y'z'} dt \; .
    \end{equation}
\end{enumerate}
%\rem{On est d'accord que le terme ci-dessus devrait plutot etre $\sG_{k'kk''}^{yxz}$ comme ci-dessous?}
Summing up all these contributions, we get
\begin{equation}
     T^{kk'}_{(x,y) \to (x',y') } = \Lambda_{xy}^{x'y'} 
     + \delta_{yy'} \left[ \Lambda_{x}^{x'} 
     + \sum_{k'' z z'}\frac{\left(\# \sG_{kk'k''}^{xyz}\right)}{\left(\# \sG_{kk'}^{xy}\right)} \Lambda_{xz}^{x'z'}\right]
     + \delta_{xx'} \left[ \Lambda_{y}^{y'} 
     + \sum_{k'' z z'}\frac{\left(\# \sG_{k'kk''}^{yxz}\right)}{\left(\# \sG_{k'k}^{yx}\right)} \Lambda_{yz}^{y'z'}\right] \; .
\end{equation}
For our SIR model on networks,  this leads for instance to
\begin{equation}
     T^{kk'}_{(s,i) \to (i,i) } = \lambda_{kk'}(t) 
     + \sum_{k''} \lambda_{kk''}(t) \frac{\left(\# \sG_{kk'k''}^{sii}\right)}{\left(\# \sG_{kk'}^{si}\right)}  \; .
\end{equation}

\subsection{The Pairwise Approximation}
\label{pairwise}

As always, we need the $\left( \# \sG_{kk'}^{xy} \right)$ to compute the evolution of the  $\left( \# \sG^x_k\right)$, but we need the $\left(\# \sG_{kk'k''}^{xyz}\right)$ to compute the evolution of the $\left( \# \sG_{kk'}^{xy} \right)$.  We can move forward if we assume that the three-body correlations are negligible. For Markovian networks, loops are rare, in the sense that the probability that a given node $\alpha$ belongs to a loop of (fixed) finite length $L$ goes to zero as the size of the network goes to infinity, and therefore this is a very good approximation.
Within this (pairwise) approximation, given a node $\alpha$, the degree and state of two of its neighbors $\beta$ and $\gamma$ are uncorrelated, so that the joint probability of having $\beta$ of degree $k'$ and state $y$ and $\gamma$ of degree $k''$ and state $z$ is essentially the product of the two probabilities $G^{xy}_{kk'}$ and $G^{xz}_{kk''}$. More precisely, we get
%\begin{equation}
%    \sG_{kk'k''}^{xyz} = \{ (\alpha,\beta,\gamma) \;/\; \alpha \in \sD_k \;\&\; (\alpha,\beta) \in \sG^{xy}_{kk'} \;\&\; (\alpha,\gamma\neq\beta) \in \sG^{xz}_{kk''}  \} \; ,
%\end{equation}
\begin{equation}
\label{Gkkkxxx}
    \left( \# \sG_{kk'k''}^{xyz}\right) \simeq \underbrace{N_k X_k}_{\# \sG^{x}_{k}} \; \underbrace{k}_{\# \mbox{ of } \beta} \overbrace{G^{xy}_{kk'}}^{\mbox{conditional proba}} \underbrace{(k-1)}_{\# \mbox{ of } \gamma} \overbrace{G^{xz}_{kk''}}^{\mbox{conditional proba}}   \; ,
\end{equation}
%\rem{$N_k X_k$ est plutot $\# \sG^{x}_{k}$, non?}
The coefficient $k(k-1)$ in \eqref{Gkkkxxx} corresponds to the choice of the two neighbors of node $\alpha$, taking into account the fact that not only the edges but also the wedges are oriented (see Section \ref{settings_defs}).  
From Eq.~\eqref{eq:Gdef} we have $ {\# \sG^{xy}_{kk'}} = G^{xy}_{kk'}  { k N_k X_k} = k' N_{k'} X_{k'} G^{yx}_{k'k}$, hence
\begin{equation}
     T^{kk'}_{(x,y) \to (x',y') } = \Lambda_{xy}^{x'y'} 
     + \delta_{yy'} \left[ \Lambda_{x}^{x'} 
     + \sum_{k'' z z'} (k-1) G_{kk''}^{xz} \Lambda_{xz}^{x'z'}\right]
     + \delta_{xx'} \left[ \Lambda_{y}^{y'} 
     + \sum_{k'' z z'} (k'-1)G_{k'k''}^{yz} \Lambda_{yz}^{y'z'}\right] \; .
\end{equation}
For our SIR model on networks, this gives in particular 
\begin{equation}
     T^{kk'}_{(s,x) \to (i,x)} = \lambda_{kk'} \, \delta_{x,i} 
     + (k-1) \sum_{k''} \lambda_{kk''} G_{kk''}^{si} \; ,
\end{equation}
which %reintroducing the $(k,k')$ dependence of $\lambda(t)$ 
is Eq.~(1c) of the main text.

\subsection{Getting to Eqs.~(2)--(3) of the main text}

Equations (2) of the main text are a direct consequence of the definition of the one-point dressed rates $T^{k}_{x \to x'}$.  Indeed,
\begin{equation}
    (\#\sG^x_k) (t+dt) - (\#\sG^x_k) (t) = 
    \sum_{x' \neq x} \sT^{k}_{(x';t) \to (x;t+dt)} 
    - \sum_{x' \neq x}\sT^{k}_{(x;t) \to (x';t+dt) }  \; ,
\end{equation}
which dividing both sides by $N_k dt$ gives
\begin{equation} \label{SI:Xdot}
\dot{X_k} = \sum_{x'} \left( X'_k T^k_{x' \to x} - X_k T^k_{x \to x'} \right) \; .
\end{equation}
Since for the SIR model the only non-zero  one-point dressed rate are $T^k_{s \to i}$ and $T^k_{i \to r}$, this readily gives Eq.~(2).

In the same way,
\[
%\int_t^{t+dt} 
\dot{\left (\# \sG_{kk'}^{xy} \right)} dt = 
 \sum_{(x'y') \neq (x,y) } \left(\# \sT^{kk'}_{(x',y';t) \to (x,y;t+dt) } 
 -\# \sT^{kk'}_{(x,y;t) \to (x',y';t+dt) } \right)
\]
and thus, removing the constant factor $k N_k dt$ which appears on both sides, we get
\begin{equation}
   \frac{d}{dt}(X_k G_{kk'}^{xy})  = 
   \sum_{(x'y') \neq (x,y)} \left( X'_k \, G_{kk'}^{x'y'} \, T^{kk'}_{(x',y') \to (x,y) }
 - X_k \, G_{kk'}^{xy}  T^{kk'}_{(x,y) \to (x',y') } \right)
   \; ,
\end{equation}   
which is Eq.~(3) of the main text.

\section{Validity of our batching procedure with the pairwise approximation}
\begin{figure}[h]
\centering
\includegraphics[scale=0.4]{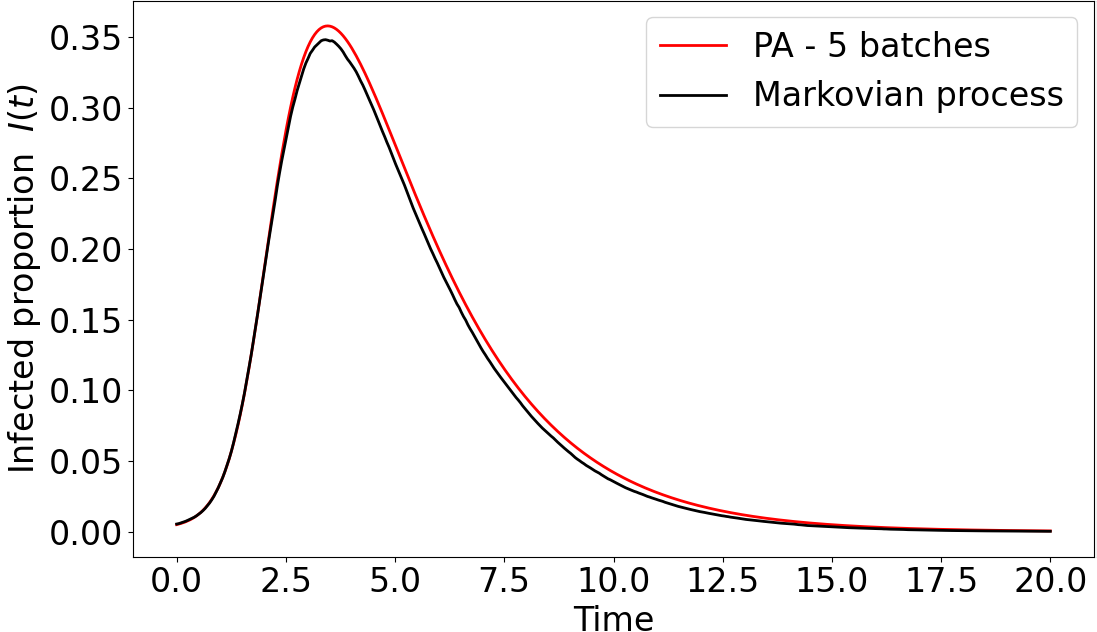}
\caption{Evolution of total infected proportion over time on a heterogeneous network. Red line: results provided by the pairwise approximation (PA) and batching procedure ($5$ batches) applied to the realistic heterogeneous network utilized in the second part of the main text. Black line: average Markovian process over $n_{\text{it}} = 10$ iterations, with $N=15\,000$ nodes. Parameters of the epidemic are $\beta = 4, \gamma=1, I_0=5.10^{-3}$.}
\label{fig:comparison_approximations_batching_procedure}
\end{figure}

The pairwise approximation described above, combined with the batching procedure applied with $5$ bins in the main text, provides a highly accurate representation of the network dynamics we aim to reproduce. This is illustrated in Fig.~\ref{fig:comparison_approximations_batching_procedure}. The small discrepancies observed do not significantly affect the general observations or the conclusions drawn regarding the Nash equilibrium on such networks.
